\documentclass[12pt,pre,aps,preprint]{revtex4}
\usepackage{psfig}

\begin{document}
\title{Voronoi and Voids Statistics for Super-homogeneous Point Processes}
\author{Andrea Gabrielli$^{1}$, Salvatore Torquato$^{2}$}
\affiliation{$^{1}$``E. Fermi'' Center for Research and Studies,
Via Panisperna 89A, Compendio del Viminale, 00184 Rome, Italy.\\
$^{2}$Department of Chemistry and Materials 
Institute, Princeton University, Princeton, NJ 08544.}

\date{\today}

\begin{abstract}

We study the Voronoi and void statistics of super-homogeneous
(or hyperuniform) point patterns in which the infinite-wavelength
density fluctuations vanish. Super-homogeneous
or hyperuniform point patterns arise in one-component plasmas,
primordial density fluctuations in the Universe, and in jammed 
hard-particle packings. 
We specifically analyze a certain one-dimensional model by studying
size fluctuations and correlations of the associated Voronoi cells. 
We  derive exact results for the complete joint statistics of the 
size of two Voronoi cells. We also provide a sum rule that the 
correlation matrix for the Voronoi cells must obey in any space dimension.
In contrast to the conventional picture of super-homogeneous
systems, we show that infinitely large Voronoi cells
or voids can exist in super-homogeneous point processes in any dimension. 
 We also present two heuristic
 conditions to identify and classify any
 super-homogeneous point process in terms of the asymptotic behavior of
 the void size distribution.

\end{abstract}
\maketitle

\section{Introduction}

Point patterns are ubiquitous in Nature. Examples include those
defined by the coordinates of the particles in a many-particle system,
such as the molecules of a liquid or crystal, stars of a galaxy, or
trees in a forest. Understanding how the number of points fluctuates
at a given length scale reveals important structural information about
the point pattern. Such local density fluctuations have been studied
for a variety of physical systems, including one-component plasmas
\cite{Ma80}, molecular liquids \cite{Tr98}, and the large-scale
structure of the universe \cite{CDM}.

Point patterns in which the infinite-wavelength density fluctuations
vanish, referred to as {\em super-homogeneous} \cite{CDM} or {\em
hyperuniform} \cite{TS2003}, are of particular interest to us in the
present paper. Regular lattices of points in space are the simplest
examples of super-homogeneous point patterns, but such point processes
are neither statistically spatially stationary (homogeneous) nor
isotropic. Stochastic super-homogeneous point processes and
fluctuations have been demonstrated to be very important in 
a variety of physical
contexts, including the study of one component plasmas
\cite{OCP1}, the evolution of primordial matter density fluctuations in
cosmology \cite{CDM}, and the structural properties of jammed
configurations of hard spheres systems \cite{TS2003}. It is considerably more difficult to construct point
patterns that are statistically stationary and isotropic, although
some examples have been identified \cite{Ma80,CDM,TS2003}.  
In order to shed light
on this problem, our general goal is to understand the statistics of
the underlying Voronoi cells associated with the points of stationary
and isotropic super-homogeneous point processes in arbitrary space
dimension $d$.  A Voronoi cell associated with a given point consists
of the region of space closer to this point than to any other point
\cite{torquato-book}.

A first step toward the stated goal is to start by examining
stationary super-homogeneous point processes in one dimension (where
isotropy is not an issue).  Specifically, we  analyze such a
particular one-dimensional model by studying size fluctuations and
correlations of the associated Voronoi cells. We  derive exact
results for the complete joint statistics of the size of two Voronoi
cells.  It is additionally shown that infinitely large Voronoi cells
can exist in super-homogeneous point processes in one dimension. We
also provide a sum rule that the correlation matrix for the Voronoi
cells must obey in any space dimension.

\section{Preliminaries}

Before discussing the details of the model, we recall some general
definitions of basic quantities that are used to statistically
characterize point processes (for rigorous definitions and analysis
see \cite{daley}).

A single realization of a point process is completely determined by
the stochastic {\em microscopic density} function $\hat n(x)$, which
in $d$ dimension, can be expressed as
\begin{equation}
\hat n({\bf x})=\sum_i \delta({\bf x}-{\bf x}_i)\,,
\label{eq1}
\end{equation}
where $\delta({\bf x})$ is the usual $d-$dimensional Dirac delta
function, $\bf{x}_i$ is the position of the $i^{th}$ point in the system
and the sum is over all of the points. The microscopic density has the
following integral property:
\[\int_{\cal V} d^dx\, \hat n({\bf x})= N[{\cal V}]\,,\]
where ${\cal V}$ is any measurable set of the space (i.e., the
one-dimensional line in the one-dimensional case of interest) and
$N[{\cal V}]$ is the number of points (particles centers) contained in
that set.

The statistics of a point process is completely determined by the
infinite set of correlation functions:
\[I_m({\bf x}_1,...,{\bf x}_m)\equiv\left< \hat n({\bf x}_1)...
\hat n({\bf x}_m)\right>\,,\]
for any integer $m\ge 1$, and where $\left<...\right>$ indicates the
ensemble average over all the possible realizations of the point
process. For definitions of more general correlation functions see
Ref. \cite{torquato-book}. Clearly, all of the functions $I_m$ are
invariant under any permutations of the variables ${\bf x}_1,...,{\bf
x}_m$.  For $n=1$, we have that
\[I_1({\bf x})\equiv n({\bf x}) =\left<\hat n({\bf x})\right>\]
gives the {\em local average density} of points at the spatial
position ${\bf x}$ and characterize completely all the one-point
statistical properties of the system.  However, very often a constant
{\em global} average density is also evaluated through a volume
average:
\begin{equation}
n_0=\lim_{V\rightarrow +\infty} \frac{1}{V}
\int_V d^dx \, \hat n({\bf x})\,,
\label{vol-av}
\end{equation}
which gives the average density of particles in the system as a whole,
and where $V$ is for example a spherical volume.. Systems in which
{\em volume} averages as in Eq.~(\ref{vol-av}) are equal to the relative
{\em ensemble} averages are called {\em ergodic systems}.  The
quantity $I_2({\bf x},{\bf y})d^dx\,d^dy$ gives the joint {\em a
priori} probability of finding a point in the volume element $d^dx$
around ${\bf x}$ and at the same time another in the element $d^dy$
around ${\bf y}$.  It is the most commonly used function to study the
correlation properties of an empirical particle distribution.

If all the $I_m({\bf x}_1,...,{\bf x}_m)$ are invariant under a
constant translation of all the points, i.e. if $I_m({\bf
x}_1,...,{\bf x}_m)=I_m({\bf x}_1+{\bf x}_0,...,{\bf x}_m+{\bf x}_0)$
for any ${\bf x}_0$ and $m$, the stochastic point process is said to
be {\em spatially statistically stationary} (or {\em statistically
homogeneous}).  In most of what follows we will limit our
considerations to this class of point process.  In such instances,
$n({\bf x})=n_0>0$ (the condition $>0$ excludes fractal point
distributions) does not depend on ${\bf x}$, and $I_2({\bf x},{\bf
y})=I_2({\bf x}-{\bf y})$ depends only on the displacement vector.  If
moreover the system is statistically isotropic $I_2$ depends only on
the scalar distance $|{\bf x}-{\bf y}|$.

A $d-$dimensional point-process is said to be {\em ergodic}
if, for any function $F[\hat n({\bf x}_1),\hat n({\bf x}_2),...,\hat
n({\bf x}_l)]$ of the microscopic density $\hat n({\bf x})$ in the
arbitrary points ${\bf x}_1,{\bf x}_2,...,{\bf x}_l$ (where $l$ is
finite but arbitrary), the following relation holds:
\begin{equation}
\lim_{V\rightarrow +\infty}\frac{1}{V}\int_V d^dx_0\,
F[\hat n({\bf x}_0+{\bf x}_1),
\hat n({\bf x}_0+{\bf x}_2),...,\hat n({\bf x}_0+{\bf x}_l)]=
\left<F[\hat n({\bf x}_1),\hat n({\bf x}_2),...,\hat n({\bf x}_l)]\right>\,.
\label{ergo}
\end{equation}
It is clear from Eq.~(\ref{ergo}) that spatial stationarity is a
necessary condition for ergodicity \cite{torquato-book,gnedenko,gardiner}.
Ergodicity is often supposed {\em apriori} as a valid working
hypothesis in the analysis of (spatially or temporally) stationary
stochastic processes.

In order to measure the density-fluctuation correlations $\delta\hat
n({\bf x})=\hat n({\bf x})-n_0$ between two different points in a
statistically stationary point process, the {\em covariance} function
(also called {\em reduced} two-point correlation function) $C({\bf
x})$ is introduced via
\begin{equation}
C({\bf x})=\left<\delta\hat n({\bf x}_0)\delta\hat n({\bf x}_0+{\bf x})\right>=
I_2({\bf x})-n_0^2\,.
\label{con-cf}
\end{equation}
It is simple to show, from (\ref{eq1}) that $C({\bf x})$ can be
written as
\[C({\bf x})=n_0\delta({\bf x})+n_0^2h({\bf x})\,,\]
where $n_0\delta({\bf x})$ is the {\em diagonal} part of $C({\bf x})$
present in any stochastic point process independently of the
correlations between different spatial points and due only to the {\em
discrete} nature of the {\em massive} point-particle distribution,
while $n_0^2h({\bf x})$, meaningful for $x\ne 0$, is the {\em
non-diagonal} part characterizing the real correlation between
different points and vanishing for $x=|{\bf x}|\rightarrow \infty$.
The function $h({\bf x})$ is referred to as the {\it total correlation
function} in the theory of liquids \cite{Ha86}.

Another important quantity, characterizing the relative weight of each
Fourier mode to a realization of the stochastic point process, is the
so-called {\em power spectrum} $s({\bf k})$ (proportional to the
so-called structure factor \cite{footnote1} and called also Bartlett
spectrum \cite{bartlett63}), which is defined by
\begin{equation}
s({\bf k})=\lim_{L\rightarrow+\infty}
\left<|\delta_n({\bf k};L)|^2\right>\,,
\label{ps-gen}
\end{equation}
where 
\[\delta_n({\bf k};L)=\frac{1}{L^{d/2}}\int_{-\frac{L}{2}}^{\frac{L}{2}}\!...\!
\int_{-\frac{L}{2}}^{\frac{L}{2}}
d^dx\,\delta\hat n({\bf x})\, e^{-i{\bf k}\cdot{\bf x}}\] is the
Fourier element of the density contrast $\delta\hat n({\bf x})$ in a
cubic volume of size $L$.  It is simple to show that if the point
process is spatially stationary then $s({\bf k})$ is simply the Fourier
transform of $C({\bf x})$:
\[s({\bf k})=n_0+ n_0^2 \int d^dx\,h({\bf x})\,e^{-i{\bf k}\cdot{\bf x}}=n_0+
n_0^2\hat h({\bf k})\,,\]
where $\hat h({\bf k})$ is the Fourier transform in the
infinite volume of $h({\bf x})$.
This result implies the so-called Wiener-Khinchtine
theorem \cite{torquato-book,Cr93}, which states that the covariance 
function of a stationary point process has a
positive Fourier transform converging to $n_0$ for sufficiently
large $k$ and integrable around $k=0$.

Finally if the system is also statistically isotropic also $s({\bf
k})$ depends only on $k=|{\bf k}|$.

\section{Super-Homogeneous (Hyperuniform) Point Processes}

Here we briefly review definitions and basic properties of
super-homogeneous (or hyperuniform) point processes.  Given a spatially
stationary point process in $d$ dimensions, we can define the variance
in the number of points in a sphere ${\Omega}(R)$ of radius $R$ (the
origin of the sphere is arbitrary because of the spatial stationarity)
as
\begin{equation}
\sigma^2(R)=\left<N^2(R)\right>-\left<N(R)\right>^2\,,
\label{sup1}
\end{equation}
where 
\[N(R)=\int_{{\Omega}(R)} d^d x\,\hat n({\bf x})\,,\]
is the number of point in the sphere ${\Omega}(R)$, which
is a stochastic function.

It is simple to show that Eq.~(\ref{sup1}) can be expressed in terms
of the covariance function $C({\bf x})$ as
follows:
\begin{equation}
\sigma^2(R)=\int_{{\Omega}(R)}\int_{{\Omega}(R)}
d^dx\,d^dy\,C({\bf x}-{\bf y})\,
\label{sup2}
\end{equation}
Equivalently, we can express the same quantity in terms of the power
spectrum $s({\bf k})$ \cite{TS2003} in the following way
\begin{equation}
\sigma^2(R)=\frac{1}{(2\pi)^d}\int d^dk |w({\bf k}; R)|^2 s({\bf k})\,,
\label{sup2b}
\end{equation}
where the integral is over all the $k$ space, and 
\[w({\bf k}; R)=\int_{{\Omega}(R)}d^dx\,e^{-i{\bf k}\cdot{\bf x}}\]
is the so-called {\em window function}.

All stationary point processes can be classified in terms
of the scaling behavior of $\sigma^2(R)$ for large $R$ as follows
\cite{CDM}:
\begin{itemize}
\item If 
\[\int d^dx\,C({\bf x})=s(0)=A>0\,,\]
i.e., the correlations are mainly positive and short ranged,
then 
\[\sigma^2(R)\sim R^d\]
for sufficiently large $R$ (i.e., for $R$ larger than the range of
correlations). The prototypical example of this class of systems is
the so-called Poisson point process \cite{EPL2001,TS2003}, which can
be generated by randomly placing points in the space with a given
average density $n_0>0$ in an uncorrelated manner. In this case, it is
simple to show that simply $C({\bf x})=n_0\delta({\bf x})$ and $s({\bf
k})=n_0$.  For this reason we call this class of point patterns {\em
essentially Poissonian}.  This is the most common behavior for the
number fluctuations for homogeneous systems in thermal equilibrium
(e.g., an ordinary gas in equilibrium at high temperature or a liquid
away from critical points).

\item If, instead,
\[\int d^dx\,C({\bf x})=s(0)=+\infty\,,\]
with $s({\bf k})\sim k^{-\gamma}$ for sufficiently small $k$
where, for definiteness, $0<\gamma<d$, then 
\[\sigma^2(R)\sim R^{d+\gamma}\]
for sufficiently large $R$. In this case, two-point correlations are
again mainly positive but are long-ranged. This situation characterizes
order parameters of a thermodynamical system at the critical
point of a second order phase transition (e.g., the gas-liquid
transition at the critical temperature and pressure).
For this reason, we call this class {\em critical systems}.

\item Finally, if
\begin{equation}
\int d^dx\,C({\bf x})=s(0)=0\,,
\label{sup3}
\end{equation}
it is possible to show that
\begin{equation}
\sigma^2(R)\sim R^{\alpha}\,,
\label{sup3bis}
\end{equation}
with $\alpha< d$.  In particular, it is possible to show that in any
case $d-1\le \alpha <d$, i.e. $\sigma^2(R)\sim R^{d-1}$ is the minimal
scaling behavior for the number fluctuations versus $R$ for any point
process (all these considerations can be directly extended to include
also any ``genuine'' continuous stochastic mass density field
\cite{CDM}).  In this case, there is an exact balance between positive
and negative correlations in the density fluctuations in such a way to
have Eq.~(\ref{sup3}).  Therefore, infinite wavelength density fluctuations
vanish, which imparts a degree of ``order" even to stochastic
point processes that satisfy (\ref{sup3}).  At sufficiently small $k$, we have
\begin{equation}
s({\bf k})\sim k^{\gamma}\,,\
\label{sup3ter}
\end{equation}
with $\gamma>0$.  It is possible to show \cite{CDM} that $\alpha$ and
$\gamma$ are related in the following way: (i) If $0<\gamma \le 1$, we
have $\alpha=d-\gamma$; (ii) If $\gamma \ge 1$, then $\alpha=d-1$ (the
``proper" condition for superhomogeneity).  For $\gamma =1$ there will
be logarithmic corrections.  

Since for the class of systems that satisfy (\ref{sup3}) the number
fluctuations increase with the spatial scale slower than in
a large class of correlated and 
uncorrelated point process (e.g., Poisson distribution), we call them
super-homogeneous or hyperuniform point processes. 
Note that super-homogeneous point
processes are at a type of ``critical" point, but one in which the
{\it direct two-point correlation function} \cite{TS2003} rather than
the covariance $C({\bf x})$ is long-ranged.

\end{itemize}

\section{The One-Dimensional Model}

\begin{figure}
\centerline{\psfig{file=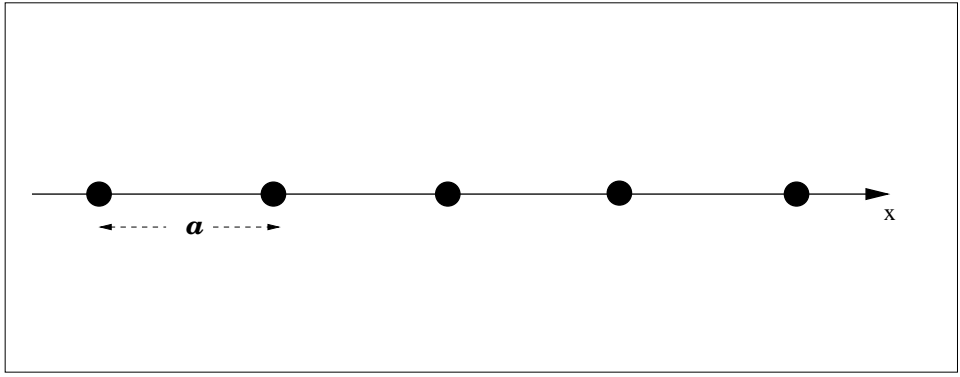,width=5.0in}}
\caption{Schematic representation of the one-dimensional lattice
with lattice constant $a$.}
\label{fig1}
\end{figure}

In order to construct a super-homogeneous point process suitable for a
complete study, we begin with a one-dimensional regular lattice of
points, i.e., a chain of point-particles with constant spatial
separation (lattice constant) $a$ (see Fig.\ref{fig1}).  The
microscopic density for such a regular point process is given by
\begin{equation}
\hat n(x)=\sum_{j=-\infty}^{+\infty} \delta(x-ja)\,,
\label{lattice1}
\end{equation}
where $a>0$ is the lattice spacing.  Clearly, such a set is not
spatially stationary, but only possesses discrete translational
invariance. However, it is the one-dimensional super-homogeneous point
process with the lowest number variance as a function of $R$
\cite{TS2003}.  The global average density of the system is simply
$n_0=1/a$.

In order to obtain a stochastic super-homogeneous one-dimensional point
process suitable for our study, we shuffle the lattice by introducing
a random displacement field. That is, we move each point from its
initial lattice position through a random displacement with a given
probability distribution, each point being displaced independently of
the others.  In practice, if the initial position of the $m^{th}$
point is $ma$, the final one will be $x_m=ma+u_m$, where $u_m$ is a
random variable extracted from the probability density function (PDF)
$p(u)$.  Note that the average density $n_0$ is not changed by the
application of the displacements, since the number of points in the
system is conserved.

It is possible to show (see Appendix \ref{app1}) that if each point of a
general initial spatial distribution is displaced from its initial
position independently of the others with a PDF $p(u)$, then the {\em
new} power spectrum $s(k)$ will be given by
\begin{equation} 
\label{disp1}
s(k)=n_0(1-|\tilde p(k)|^2)+s_{I}(k)|\tilde p(k)|^2\,,
\end{equation}
where $s_{I}(k)$ is the initial power spectrum before the
displacements and
\begin{equation}
\tilde p(k)=\int_{-\infty}^{+\infty}du \,p(u)\, e^{-iku}
\label{char-f}
\end{equation} 
is the Fourier transform of $p(u)$, i.e. the so-called {\em
characteristic function} of the random-displacement PDF (for a more
general discussion of the effect of a stochastic displacement field
with arbitrary spatial correlation on a given point process see
\cite{gabrielli1}). In general, we take $p(u)$ to be symmetric, i.e.,
$p(u)=p(-u)$.  Note that for all possible $p(u)$, we have the limit
condition $\tilde p(0)=1$ and that for small $k$ in the symmetric case
\begin{equation}
\tilde p(k)\simeq 1- Ak^\alpha
\label{disp2}
\end{equation} 
with $\alpha=2$ and $A=\frac{\overline{u^2}}{2}$ if $\overline{u^2}$
is finite, and where
$\overline{f(u)}=\int_{-\infty}^{+\infty}du\,p(u)f(u)$ means the
average over the uncorrelated displacements. For $d=1$, this is the
case if $p(u)$ decreases faster than $|u|^{-3}$ for large $|u|$.
Otherwise, if $\overline{u^2}=+\infty$, i.e., $p(u)\simeq
B|u|^{-\beta}$ for large $|u|$ with $1<\beta<3$, then $\alpha=\beta-1$
and \cite{gabrielli1}
\begin{equation}
A=2B\int_0^{+\infty} dx x^{-\beta}(1-\cos x)\,,
\label{disp3}
\end{equation} 
where $B$ is a positive constant.

In the case of a lattice, it is well known \cite{CDM} and simple 
to show that 
\[s_{I}(k)=\frac{2\pi}{a^2}\sum_{m\ne 0}
\delta\left(k-\frac{2\pi m}{a}\right)\,,\]
where the sum is over all of the integers $m$, except $m=0$.
Therefore, from Eq.~(\ref{disp1}), the power spectrum of the ``shuffled''
lattice is
\begin{equation} 
\label{disp1bis}
s(k)=\frac{1-|\tilde p(k)|^2}{a}+\frac{2\pi}{a^2}\sum_{m\ne 0}
\delta\left(k-\frac{2\pi m}{a}\right)\left|\tilde p
\left(\frac{2\pi m}{a}\right)\right|^2\,.
\end{equation}
Recall that superhomogeneity (or hyperuniformity) of the point process
is given by only the behavior of $s(k)$ in the vicinity of $k=0$.
Therefore, since in the {\em first Brillouin zone} the power spectrum
of a lattice is identically zero (i.e. the first Bragg peaks are at
$|k|=2\pi/a$), the small $k$ behavior of $s(k)$ is determined only by
that of $\tilde p(k)$.  In particular, for $|k|< 2\pi/a$ and $|k|\ll
(1/A)=^{1/\alpha}$ [cf. Eqs.~(\ref{disp2}) and (\ref{disp3})] we have
from the discussion above that
\begin{equation}
s(k)= \frac{2Ak^\alpha}{a}\;\;\mbox{with}\left\{
\begin{array}{ll}
\alpha=2\mbox{  and  }A=\frac{\overline{u^2}}{2}\,,
&\mbox{ if }\overline{u^2}<+\infty\\
\alpha=\beta-1\mbox{  and  }A \mbox{ from Eq.~(\ref{disp3}), }&\mbox{ if }
\overline{u^2}=+\infty
\end{array}
\right.
\label{disp4}
\end{equation}
which always satisfies the superhomogeneity condition $\alpha>0$.  In
particular, for $\beta> 2$ we have $1< \alpha\le 2$, and the
condition of {\em minimal} mass fluctuations-length scaling for point
process in $d$ dimensions (i.e. $\sigma^2(R)\sim R^{d-1}$) is
satisfied (for $\beta=2$ there are logarithmic corrections in $L$).

In the case in which each point is completely randomly
displaced inside its own unit cell, i.e.,
\[p(u)=\frac{\theta\left(\frac{a}{2}-|u|\right)}{a}\,,\]
where $\theta(x)$ is the usual Heaviside step function,
the final point distribution is not only super-homogeneous,
but also completely statistically stationary (i.e., with a complete
statistical translational invariance), even though the original
lattice array was not. 

\begin{figure}
\centerline{\psfig{file=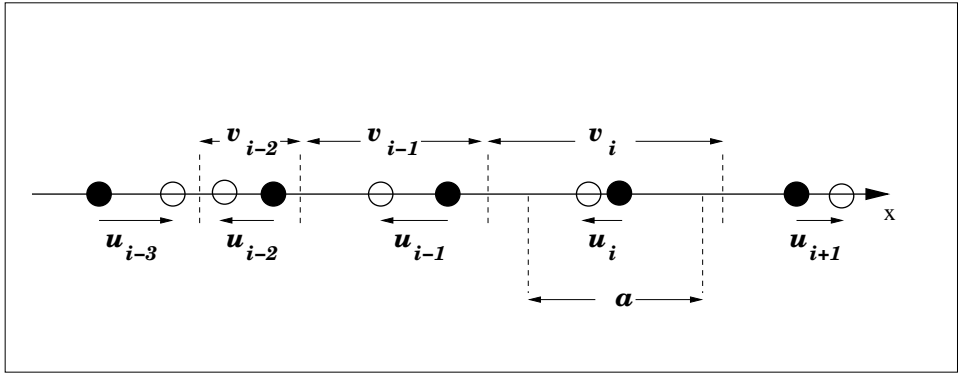,width=5.0in}}
\caption{``Shuffled'' lattice with the PDF of the uncorrelated 
displacements
$p(u)=\frac{\theta\left(\frac{\Delta}{2}-|u|\right)}{\Delta}$ with
$\Delta\le a$. The filled circles represent the initial lattice
configuration (i.e., a lattice with a lattice constant $a$), while the
empty circles are the new positions of the points after the
displacements $u_i$. The quantity $v_i$ is the size of the final
Voronoi cell of the point initially at the lattice position $a\cdot
i$.  }
\label{fig2}
\end{figure}

We first analyze the behavior of the fluctuations 
associated with the volumes of the Voronoi cells 
in the simple case in which (see Fig.\ref{fig2})
\begin{equation}
p(u)=\frac{\theta\left(\frac{\Delta}{2}-|u|\right)}{\Delta}
\mbox{  with }\Delta\le a\,.
\label{disp5}
\end{equation}

The statistics of the Voronoi cells are relatively simple because no
point is allowed to move into the unit cell centered at the initial
position of another point. In what follows, starting from the results
for this model, we will extend some of the results to the most general
class of super-homogeneous point processes in any dimension.

\section{Voronoi-Cell Statistics}  

As stated above, we start from a regular lattice of points with
microscopic density given by Eq.~(\ref{lattice1}), and displace each
point independently of the others by applying to it a displacement
whose PDF $p(u)$ is given by Eq.~(\ref{disp5}).  Taking
the Fourier transform of this PDF yields the
characteristic function $\tilde p(k)$ to be exactly given by
\[\tilde p(k)=\frac{\sin(\frac{k\Delta}{2})}{\frac{k\Delta}{2}}\,.\]
Consequently, applying Eq.~(\ref{disp1bis}) we obtain
\begin{eqnarray}
\label{voro1}
s(k)=&&\frac{1}{a}\left[1-\left(\frac{2}{k\Delta}\sin\left(\frac{k\Delta}{2}
\right)\right)^2\right]+\\
&&\frac{2\pi}{a^2}\sum_{m\ne 0}
\delta\left(k-\frac{2\pi m}{a}\right)\left(\frac{a}{m\pi\Delta}
\sin\left(\frac{m\pi\Delta}{a}\right)\right)^2\,.\nonumber
\end{eqnarray}
We can verify directly that, since $\sin(m\pi)=0$ for any integer $m$,
only if $\Delta=a$ the contribution to Eq.~(\ref{voro1}) coming from
the Bragg peaks of the underlying lattice structure completely
vanishes. In fact, it is the only case in which the point process is
fully translationally invariant.

\begin{figure}
\centerline{\psfig{file=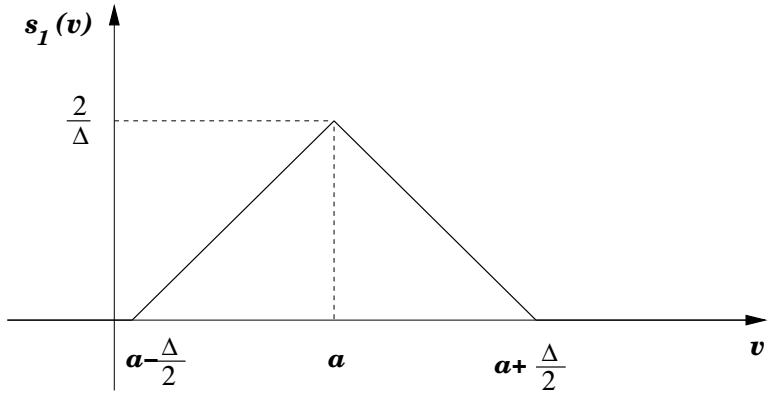,width=4.0in}}
\caption{Representation of the one-cell size PDF $f_1(v)$ for our
model.
}
\label{fig3}
\end{figure}

We can now proceed to the evaluation of the statistics of the Voronoi
cells. For a point process in any dimension, the Voronoi cell
associated with a given point consists of the region of space closer
to this point than to any other point.  The collection of all of the
Voronoi cells that tiles the space is referred to as a Voronoi {\it
tessellation}.  Clearly, in the initial lattice configuration, the
Voronoi cell associated with each point coincides with the unit cell
of size (length) $a$ around each point.  According to
Eq.~(\ref{disp5}), a randomly displaced point that was at the original
lattice position $ja$ (integer $j$) remains within its original unit
cell. Consequently, we will always refer to this as point $j$. The
size of its new Voronoi cell $v_j$ will be given, by definition, by
the size of the line segment that joins the point that lies exactly
midway between the points $j+1$ and $j$ and the point lies exactly
midway between the points $j$ and $j-1$, i.e.,
\begin{equation}
v_j=a+\frac{u_{j+1}-u_{j-1}}{2}\,,
\label{voro1b}
\end{equation} 
where $u_j$ is the displacement applied to point $j$.  The PDF
$f_1(v)$ characterizing the 
size of the single Voronoi cell is formally given by
\begin{equation}
f_1(v)=\int\!\!\int_{-\infty}^{+\infty}
dx\,dy\, p(x)p(y)\delta\left(v-a-\frac{x-y}{2}\right)\,.
\label{voro2}
\end{equation}
Use of  Eq.~(\ref{disp5}) yields (see Fig.\ref{fig3})
\begin{equation}
f_1(v)=\frac{2}{\Delta^2}
\left\{
\begin{array}{lll}
0&\mbox{if}&|v-a|\ge \frac{\Delta}{2}\\
2(v-a)+\Delta&\mbox{if}& -\frac{\Delta}{2}\le v-a \le 0\\
-2(v-a)+\Delta&\mbox{if}& 0\le v-a \le \frac{\Delta}{2}
\end{array}
\right.
\label{voro3}
\end{equation}
Let $\overline{(...)}$ denote the average over the realizations of the
displacement field.  Since we start from a deterministic point
distribution (i.e., a lattice), this average is equivalent to the
ensemble average over the final point process. In general, when
also the initial state is a realization of a stochastic point process,
the ensemble average over the final configurations by the double
average $\left<\overline{(...)}\right>$ must be taken, where
$\left<...\right>$ is the average over the realizations of the initial
point process, and $\overline{(...)}$ is the average over the
displacements {\em conditioned} to the initial configuration.  If the
realization of the displacement field, seen as a continuous stochastic
field with a value $u({\bf x})$ in each spatial point, is independent
of the realization of the initial point distribution, the order of the
two averages is totally arbitrary. It is only under this hypothesis
that Eq.~(\ref{disp1}) is valid.

Clearly, the average size of a Voronoi cell is given by
\[\overline{v}\equiv\int_0^\infty dv\, v\,f_1(v) =a\,.\]
The variance of the size of the Voronoi cell is given by
\[\overline{v^2}-\overline{v}^2=\frac{\Delta^2}{24}\,.\]
Note that, as only finite up to $\Delta/2$ jumps are permitted, only
finite fluctuations for $v$ are possible.  The interesting question of
whether infinitely large cell-size fluctuations are permitted in a
super-homogeneous point process will be tackled in the next section
together with other important aspects of Voronoi cells 
fluctuations.

In the rest of this section, we analyze the joint probability
distribution of two different Voronoi cells. In particular,
we  find an important ``conservation law'' for cell-cell correlations.  

\begin{figure}
\centerline{\psfig{file=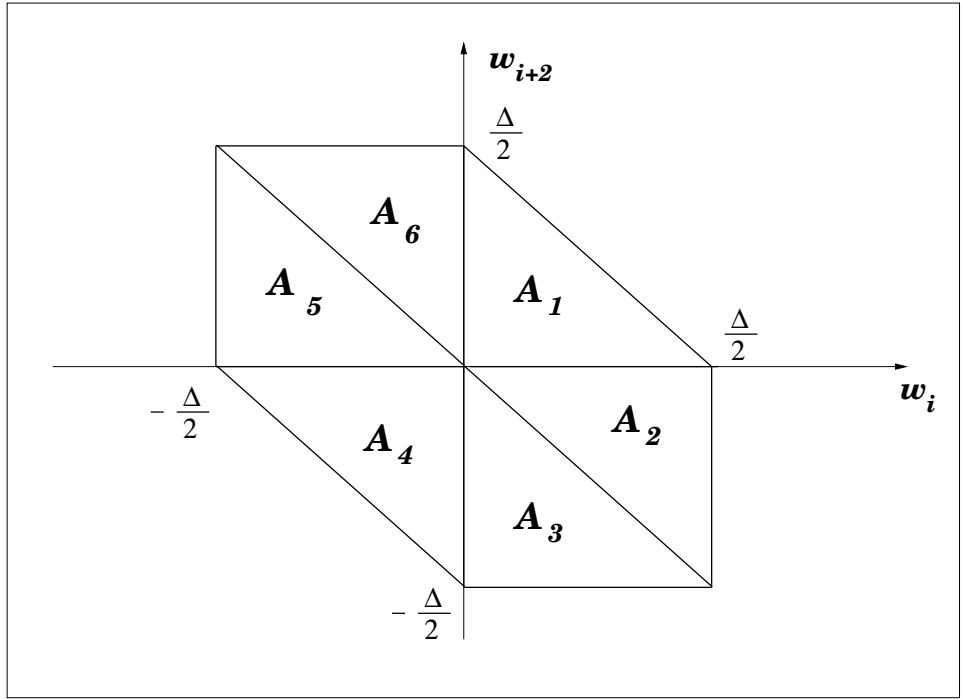,width=5.0in}}
\caption{Regions of the plane$(w_i=v_i-a,w_{i+2}=v_{i+2}-a)$ where
the joint PDF $f_2(v_i,v_{i+2})\ne 0$.}
\label{fig4}
\end{figure}

In order to find
the two-cell joint PDF $f_2(v_i,v_j)$, it is important to note that
in light of Eq.~(\ref{voro1b}) $v_i$ and $v_j$ are two dependent
variables only if $|i-j|=2$.
This means that for $|i-j|\ne 2$, we have
\[f_2(v_i,v_j)=f_1(v_i)f_1(v_j)\,.\]
For $j=i+2$, the PDF  $f_2(v_i,v_{i+2})$ will be given by the integral
\begin{eqnarray}
f_2(v_i,v_{i+2})=&&\int\int\int_{-\infty}^{+\infty}du_{i-1}du_{i+1}du_{i+3}
\,p(u_{i-1})p(u_{i+1})p(u_{i+3})\cdot\nonumber\\
&&\delta\left(v_i-a-\frac{u_{i+1}-u_{i-1}}{2}\right)
\delta\left(v_{i+2}-a-\frac{u_{i+3}-u_{i+1}}{2}\right)\,,
\label{voro4}
\end{eqnarray}
where $p(u)$ is still given by Eq.~(\ref{disp5}).
By performing explicitly the calculations and calling $w_j=v_j-a$ 
for all $j$, it is simple to show that
\begin{equation}
f_2(v_i,v_{i+2})=\frac{4}{\Delta^3}\left\{ 
\begin{array}{lll} 
\Delta-2(w_i+w_{i+2})&\mbox{in}& A_1\\
\Delta-2w_i&\mbox{in}& A_2\\
\Delta-2w_{i+2}&\mbox{in}& A_3\\
\Delta+2(w_i+w_{i+2})&\mbox{in}& A_4\\
\Delta+2w_i&\mbox{in}& A_5\\
\Delta-2w_{i+2}&\mbox{in}& A_6\\
0&\mbox{elsewhere}&
\end{array} 
\right.  
\label{voro5} 
\end{equation} 
where (see Fig.\ref{fig4}) the $A_i$ are the joint conditions:
\begin{eqnarray*}
 A_1 &=& \{w_i\ge 0
\: \mbox{and} \;0\le w_{i+2}\le -w_i+\Delta/2\}\\
 A_2 &=& \{0\le w_i\le \Delta/2 \: \mbox{and} \;
-w_i\le w_{i+2}\le 0\} \\
 A_3 &=& \{w_i\ge 0 \: \mbox{and} \;
-\Delta/2\le w_{i+2}\le -w_i\} \\
 A_4 &=& \{w_i\le 0  \: \mbox{and} \;
-w_i-\Delta/2\le w_{i+2}\le 0\} \\
 A_5 &=& \{-\Delta/2\le w_i\le 0 \: \mbox{and} \;
0\le w_{i+2}\le -w_i\} \\
 A_6 &=& \{w_i\le 0 \: \mbox{and} \;
-w_i\le w_{i+2}\le \Delta/2\}.
\end{eqnarray*}

The most basic and important quantity characterizing 
correlations between the size of different Voronoi cells is given by the
correlation matrix $C_{ij}$ defined by
\[C_{ij}=\overline{(v_i-a)(v_j-a)}\,,\]
where in this case the average is taken by using $f_2(v_i,v_j)$
[cf. Eq.~(\ref{voro5})].
Clearly, $C_{ii}=\overline{v^2}-\overline{v}^2$.
By direct calculation we have
\begin{equation}
C_{ij}=\left\{
\begin{array}{lll}
\frac{\Delta^2}{24}&\mbox{for}&i=j\\
-\frac{\Delta^2}{48}&\mbox{for}&i=j\pm 2\\
0&\mbox{for}&i\ne j,j\pm 2\\
\end{array}
\right.
\label{voro6}
\end{equation}
We see that different Voronoi cells are either anti-correlated or
uncorrelated in such a way that
\begin{equation}
\sum_{j=-\infty}^{+\infty}C_{ij}=0\,,
\label{voro7}
\end{equation}
i.e., positive and negative correlations must balance so that the sum
of $C_{ij}$ over $j$ is exactly zero. Because of the strong
resemblance with the basic property Eq.~(\ref{sup3}) of all the
super-homogeneous point processes in arbitrary $d$ dimensions, we
expect that Eq.~(\ref{voro7}) is a general property of all
super-homogeneous point processes in any dimension.

To show that this expectation is indeed true, consider a spatially
stationary super-homogeneous point process in $d$ dimensions with
average density of points $n_0>0$. For such a point process, we know
that the variance in the number of points $N(R)$ in a sphere of radius
$R$ for sufficiently large $R$ satisfies the relation
\begin{equation}
\left<N^2(R)\right>-\left<N(R)\right>^2\sim R^{\alpha}\mbox{  with }
d-1\le \alpha <d
\label{voro8}
\end{equation}
We will focus our attention on a given sufficiently large subset
${\cal S}$ of volume $V$ (e.g. a sphere or an ellipsoid) and consider
the number of points contained within it.  The average value of this
number is $\left<N({\cal S})\right>=n_0V$.  Let us call  $v_i$ the
volume of the Voronoi cell associated with point $i$. Since the set of
point-particles is countable, we can arbitrarily label
and enumerate them. By definition, $\left<v_i\right>=1/n_0$.  

Let us now study the fluctuations of the quantity
\[U({\cal S})=\sum_{i=1}^{N({\cal S})} v_i\]
under the condition of superhomogeneity. Its precise value for a
single realization will fluctuate from its average value given by
\begin{equation}
\left<U({\cal S})\right>=\left<N({\cal S})\right>\left<v_i\right>=
V\,.
\label{voro9}
\end{equation} 
In light of Eq.~(\ref{voro8}), we can write
\begin{equation}
\left<|U({\cal S})-V|^2\right>\sim V^{\frac{\alpha}{d}}\,,
\label{voro10}
\end{equation}
where it is to notice that $(\alpha/d)<1$.
But from Eq.~(\ref{voro9}), we can rewrite 
\begin{equation}
\left<|U({\cal S})-V|^2\right>=\left<\sum_{i,j}^{1,N({\cal S})}
w_i\,w_j\right>\sim V^{\frac{\alpha}{d}} \,,
\label{voro11}
\end{equation} 
where, as in the one-dimensional case, $w_i=v_i-1/n_0$.  This equation
with $\alpha<d$ (condition of superhomogeneity), together with the
fact that $N({\cal S})$ grows proportionally to $V$ and the supposed
spatially stationarity of the stochastic point process, implies
directly that in the limit of an infinite subset $\cal S$ ,we have
\begin{equation}
\lim_{V\rightarrow+\infty} \left<\sum_{j=1}^{N({\cal S})}
w_i\,w_j\right>=\sum_j C_{ij}= 0\,,
\label{sum2}
\end{equation}
where $C_{ij}=\left<w_i\,w_j\right>$ and the last sum is extended over
all of the point $j$ of the system in the infinite volume limit.  This
result can be shown rigorously by various techniques, but it is
sufficiently self evident to avoid having to present the mathematical
details.  This result is valid for any Voronoi cell $i$.  In fact,
while the matrix $C_{ij}=\left<w_i\,w_j\right>$ depends on the way we
have enumerated the points, in the case of spatially stationary point
process, the sum $\sum_j C_{ij}$ does not depend on the enumeration.
This is a quite interesting aspect of relation (\ref{sum2}). 

Therefore, in addition to Eq.~(\ref{sup3}), we have found another
``sum rule'' that applies to all spatially stationary super-homogeneous
point processes.  To check that non-super-homogeneous point processes
do not generally satisfy Eq.~(\ref{sum2}) is a very simple task.  
In fact from Eq.~(\ref{voro11}) it is simple to see that if $\alpha\ge d$
Eq.~(\ref{sum2}) can not hold.

\section{Large Cell-Size Fluctuations in Super-Homogeneous 
Point Processes and void distribution}

In the previous sections, we analyzed the main properties
of the one- and two-point statistics of Voronoi cells for 
super-homogeneous point processes. We found an important sum rule
involving that the sum along any line or column of the Voronoi cells 
correlation vanishes for any super-homogeneous point process.  
In this section, we tackle two more important questions about
super-homogeneous point processes: (1) Can there be
infinitely large Voronoi cells, or, equivalently, infinitely large
voids, for super-homogeneous point processes? (2) Is it possible
to find a functional expression for void size distribution
linking the probability of having a void of a certain size
to the correlation properties of the super-homogeneous point process?
We will see that the answers to both questions are in the affirmative.

The first question is motivated by the following facts: 
\begin{itemize}
\item All the commonly known super-homogeneous point processes (lattices,
quasi-crystals \cite{quasi-cr}, 
the one-component plasma \cite{OCP1,OCP2}, $g_2-$invariant
processes \cite{TS2003}
etc.) in the infinite-volume limit have only finite Voronoi cells and
spherical voids; 
\item By taking Eq.~(\ref{sup2b}), for a general stochastic mass
distribution (continuous or point-like), it is possible to show
\cite{CDM} that, if $s({\bf k})\sim k^n$ at small $k$, then the wave
modes which contribute essentially to create mass (i.e. number in
point processes) fluctuations on large spatial scales $R$ satisfy:
$k\sim 1/R$ if $n<1$ (and therefore $\sigma^2(R)\sim R^{d-n}$), and
$k\sim k_0$ independent of $R$ if $n\ge 1$ (and therefore
$\sigma^2(R)\sim R^{d-1}$ for all $n\ge 1$).  In particular, $k_0$
marks the departure from the small $k$ behavior of $s({\bf k})$ to its
crossover to the large $k$ behavior; in general ``shot-noise''
behavior for a point process, and a rapid cut-off to zero for a
continuous mass distribution.  Therefore, one might surmise that, at
least in the case $n>1$, voids much larger than the inverse of this
cut-off wave mode $k_0$ are not permitted at all. This certainly is
the case for the one-dimensional model presented in the previous
section in which Voronoi cells larger than twice the original unit
cell (i.e., the inverse of the average density) are not permitted.
\end{itemize}
However, more generally, we will see here that even in the case of
Eq.~(\ref{voro8}) with $\alpha=d-1$, there are super-homogeneous point
processes for which we can find spherical (or spherical-like) voids
(and therefore Voronoi cells) that are arbitrarily large.  Moreover,
and importantly, for the case of shuffled lattices, we will derive
mathematical relations between the probability of applying large
displacements and the probability of having a void of the same size.
This will permit to us to formulate an {\em ansatz} for the
characterization of the whole class of super-homogeneous point
processes in terms of the void size distribution.

With this aim, we start again from the one-dimensional regular lattice
of the previous section with lattice constant $a=1$ and microscopic
density given by Eq.~(\ref{lattice1}). We then again apply to it again
an uncorrelated displacement field, but now we choose $p(u)$ with an
unlimited tail.  As already shown in the Eq.~(\ref{disp4}) of the
previous section, the final point process is always super-homogeneous
satisfying the condition $s(0)=0$ for all possible $p(u)$. With the
aim of simplicity but no loss of generality in the final result, we
restrict the analysis to the case in which $p(-u)=p(u)$.

Let us take the segment $[0,2R]$ (i.e. the one-dimensional sphere of
radius $R$) with $R\gg a=1$, and ask for the probability $W(R)$ that
after the application of the displacement field no point is contained
in it.  Clearly, $W(R)$ can be identified also with the probability
that a randomly chosen void has a radius larger than $R$.  Therefore,
$\omega(R)=-\frac{dW(R)}{dR}$ gives the PDF of the size (i.e. radius) 
of the voids.

Given a point-particle initially at the lattice position $m$, it is
simple to show that the probability $w_m(R)$ that after the
displacement $u$, it will outside of the segment $[0,2R]$,
independently of $m$, is
\begin{equation}
w_m(R)=1-\phi(-m)+\phi(-m+2R)\,,
\label{void1}
\end{equation}
where
\begin{equation}
\phi(x)=\int_x^{+\infty}du\,p(u)\,.
\label{void1b}
\end{equation}
Note that because $p(u)$ is integrable over all the space
\begin{equation}
\lim_{x\rightarrow +\infty}\phi(x)=0\mbox{  and  }
\lim_{x\rightarrow -\infty}\phi(x)=1
\label{void1c}
\end{equation}
in any case.
Since $W(R)$ is the probability that {\em all} of the points in the
system are outside of the segment $[0,2R]$ after the displacements,
we can write
\begin{equation}
\label{void2}
W(R)=\prod_{m=-\infty}^{+\infty}\left[1-\phi(-m)+\phi(-m+2R)\right]\,.
\end{equation}
In this equation, we can distinguish between two multiplicative
contributions by writing
\[W(R)=W_1(R)W_2(R)\,.\]
The former contribution $W_1(R)$ is given by the points initially
outside the segment $[0,2R]$, and the latter $W_2(R)$ by those
initially inside it.  We show that the large-$R$ behavior of $W(R)$ is
determined essentially by this second contribution.
\begin{enumerate}
\item Let us consider the first contribution:
\begin{equation}
W_1(R)=\prod_{m<0,m>2R}\left[1-\phi(-m)+\phi(-m+2R)\right]
\label{void3}
\end{equation}
Because of the discrete translational invariance of the initial 
configuration, $W_1(R)$ can be rewritten as
\begin{eqnarray}
\label{void4}
W_1(R)&=&\left[
\prod_{m=1}^{+\infty}\left(1-\phi(m)+\phi(m+2R)\right)\right]^2\\
&=&\exp\left[2\sum_{m=1}^{+\infty}\ln\left(1-\phi(m)+\phi(m+2R)
\right)\right]\,.\nonumber
\end{eqnarray}
For any finite value of $R$, the convergence properties of the series
\begin{equation}
\sum_{m=1}^{+\infty}\ln\left(1-\phi(m)+\phi(m+2R)\right)
\label{void4b}
\end{equation}
are given by the large-$m$ behavior of
$\ln\left(1-\phi(m)+\phi(m+2R)\right)$.  Because of Eq.~(\ref{void1c})
we can say that for sufficiently large $m$
\[\ln\left(1-\phi(m)+\phi(m+2R)\right)\simeq -\phi(m)+\phi(m+2R)\,.\]
At this point we have to distinguish two sub-cases:

(1) The PDF $p(u)$ is such that 
\[\int_{-\infty}^{+\infty}du\,|u|\,p(u)<+\infty\,.\]
In this case
\[\lim_{x\rightarrow +\infty}x\phi(x)=0\,.\]
This implies that
\[\sum_{m=1}^{+\infty}\phi(m)<+\infty\,.\]
and so $\sum_{m=1}^{+\infty}\ln (1-\phi(m))$ will do. 
Therefore, to lowest order in $1/R$, we can neglect 
$\phi(m+2R)$ with respect to $\phi(m)$ in Eq.~(\ref{void4}) and write 
\begin{equation}
W_1(R)=p_0>0\,,
\label{void5}
\end{equation}
where
$p_0=\exp\left[2\sum_{m=1}^{+\infty}\ln\left(1-\phi(m)\right)\right]$.
Corrections to Eq.~(\ref{void5}) vanish for $R\rightarrow +\infty$.

(2) If the PDF $p(u)$ is such that
\[\int_{-\infty}^{+\infty}du\,|u|\,p(u)=+\infty\,,\]
i.e. if $p(u)=B\,u^{-\beta-1}$ with $0< \beta \le 1$ for 
sufficiently large $u$, then
\[\lim_{x\rightarrow +\infty}x\phi(x)=+\infty\,.\]
This implies that 
\[\sum_{m=1}^{+\infty}\phi(m)=+\infty\,,\]
being $\phi(m)\simeq \frac{B}{\beta} m^{-\beta}$ for large $m$.
At any rate,  the convergence of Eq.~(\ref{void4b}),
for any finite $R$, is still ensured by the following observation.
In the limit $m\gg 2R$, we can write
\[\ln\left(1-\phi(m)+\phi(m+2R)\right)\simeq -\phi(m)+\phi(m+2R)\simeq
-2B\,R\,m^{-\beta-1}\,,\]
which guarantees the convergence of Eq.~(\ref{void4b}).
This implies that for large $R$, $W_1(R)$ will have this main
behavior 
\begin{equation}
W_1(R)\simeq \exp\left[-a(\beta)R^{1-\beta}\right]\,,
\label{void6}
\end{equation}
where $a(\beta)>0$. In particular, for $\beta=1$, we expect that $W_1(R)$ 
goes to zero for $R\rightarrow +\infty$ as a power law.

\item Let us now analyze the second contribution to Eq.~(\ref{void2}):
\begin{equation}
W_2(R)=\prod_{0\le m \le 2R}\left[1-\phi(-m)+\phi(-m+2R)\right]=
\prod_{0\le n \le 2R}\left[1-\phi(n-2R)+\phi(n)\right]\,,
\label{void7}
\end{equation}
where in the last step we have adopted the change of variable $n=2R-m$.
Note that $n-2R\le 0$ and that for $R\rightarrow +\infty$ with $n$ fixed
$\phi(n-2R)\rightarrow 1$. Using the symmetry property $p(-u)=p(u)$ of
the PDF of the jumps \cite{footnote2}, we can write
\[\phi(n-2R)=1-\phi(2R-n)\,.\]
Therefore, Eq.~(\ref{void7}) can be rewritten as
\begin{equation}
W_2(R)=\prod_{0\le n \le 2R}\left[\phi(2R-n)+\phi(n)\right]=
\exp\left[\sum_{0\le n \le 2R}\ln \left(\phi(2R-n)+\phi(n)\right)\right]\,,
\label{void7b}
\end{equation}
In order to evaluate $\sum_{0\le n \le
2R}\ln\left(\phi(2R-n)+\phi(n)\right)$ let us approximate the sum by 
an integral as follows:
\begin{eqnarray}
\label{void8}
\sum_{0\le n \le 2R}\ln\left(\phi(2R-n)+\phi(n)\right)&\simeq&
\int_0^{2R} dx\,\ln\left(\phi(2R-x)+\phi(x)\right)\\
&=&2 \int_0^R 
dx\,\ln\left(\phi(2R-x)+\phi(x)\right)\nonumber
\end{eqnarray}
Since $\phi(x)$ is a decreasing function of $x$, we can introduce
a further approximation by developing the $\ln$ in Taylor series to
the first order in $\phi(2R-x)/\phi(x)$ :
\begin{equation}
\sum_{0\le n \le 2R}\ln\left(\phi(2R-n)+\phi(n)\right)\simeq
 2\int_0^R 
dx\,\left[\ln\phi(x)+\frac{\phi(2R-x)}{\phi(x)}
\right]\,.
\label{void9}
\end{equation}
In general, the contribution given by the term $\phi(2R-x)/\phi(x)$
can be neglected for large $R$ with respect to the first one.
We will use this approximation to study some simple but important 
cases: (A) a power-law tailed $p(u)$, and (B) a generalized-exponential
tailed $p(u)$.

(A) Let us consider the case in which $p(u)=B\,u^{-\beta-1}$ with 
$\beta>0$ for sufficiently large $u$.
In this case for sufficiently large $R$ one obtains
\[
\int_0^{R} dx\,\ln\phi(x)\simeq -\beta R\ln (2R)\,.
\]
This implies in the same limit of large $R$
\begin{equation}
\sum_{0\le n \le 2R}\ln\left(\phi(2R-n)+\phi(n)\right)\simeq
-2\beta R\ln (2R)\,.
\label{void10}
\end{equation}
Corrections to this approximation can be neglected for large $R$
as they are of the same or lower order than $R$.
Finally, we can write 
\[W_2(R)\sim e^{-2\beta R\ln (2R)}=(2R)^{-2\beta R}\,.\]
We see that for any $\beta>0$ the quantity $W_2(R)$ decreases faster
than an exponential $\exp (-AR)$, and therefore this is the main
contribution to the behavior of $W(R)$ for large $R$, i.e.,
\[W(R)\sim W_2(R)\sim e^{-2\beta R\ln (2R)}=(2R)^{-2\beta R}\,.\]
For $\beta\rightarrow 0$, the linear corrections in $R$ to
Eq.~(\ref{void10}) dominate giving $W(R)\sim \exp (-AR)$.  This is
well understood by considering that for $\beta \rightarrow 0$ the
final configuration of the point distribution will no longer be
super-homogeneous, but Poissonian for which it is well known that the
size of voids is exponentially distributed \cite{torquato-book}.

(B) Let us consider now the case in which $p(u)\simeq c|u|^p \exp 
\left[-\left(\frac{|u|}{u_0}\right)^\alpha\right]$ at sufficiently
large $|u|$ (in particular $|u|> u_0$) with $u_0>0$, $\alpha>0$ and
any $p$.  Again, we can use the approximation
\[
\sum_{0\le n \le 2R}\ln\left(\phi(2R-n)+\phi(n)\right)\simeq
2\int_0^{R} dx\,\ln\phi(x)\,.\]
It can be shown by different techniques that for large $x\gg u_0$
\[\phi(x)\simeq c\frac{u_0^\alpha}{\alpha}x^{p-\alpha+1}
\exp 
\left[-\left(\frac{x}{u_0}\right)^\alpha\right](1+o(u_0/x))\,.\]
Therefore, for asymptotically large $R$ (i.e. $2R\gg \max[1,u_0]$), the
dominating behavior will be 
\[\sum_{0\le n \le 2R}\ln\left(\phi(2R-n)+\phi(n)\right)\simeq
-\frac{2u_0}{\alpha+1}\left(\frac{R}{u_0}\right)^{\alpha+1}\,,\]
which implies 
\[W_2(R)\sim \exp\left[-\frac{2u_0}{\alpha+1}
\left(\frac{R}{u_0}\right)^{\alpha+1}\right]\,.\]
Since in this case $W_1(R)$ approximately does not depend on $R$ for
large $R$, as in the previous case, $W_2(R)$ determines the 
behavior of $W(R)$, i.e.,
\[W(R)\sim W_2(R)\sim \exp\left[-\frac{2u_0}{\alpha+1}
\left(\frac{R}{u_0}\right)^{\alpha+1}\right]\,.\]
\end{enumerate}
From the analysis of these two examples, we expect that, if $p(u)$ is
a PDF with an unlimited tail such that (as power laws and generalized
exponentials) at large $R$
\begin{equation}
\ln p(\gamma R)= \gamma'\ln p(R)(1 + o(1))
\label{tail}
\end{equation} 
with $\gamma, \gamma'$ two positive related constants of order $1$,
then the following relation holds
\begin{equation}
W(R)\sim \exp (A R \ln p(R))\,,
\label{p-voids}
\end{equation}
where $A$ is a suitable constant depending on the average density of
points $n_0$ and on the details of $p(u)$. In fact, this result can be
generalized to any other $p(u)$ with unlimited tail \cite{footnote3}.

The extension of this result to higher dimensions, in which again a
regular lattice is perturbed by an uncorrelated displacement field
characterized by a PDF with an unlimited tail, is
straightforward when the PDF of the $d-$dimensional
displacement factorizes into a product of the PDF's of the single
components $p_d({\bf u})=\prod_{i=1}^{d}p(u_i)$.  In this case, by
following the same procedure for the one-dimensional case, one can find
that, given a cube of large size $2R$, the probability that it becomes
completely void after the application of the displacement field is
\begin{equation}
W(R)\sim \exp (A R^d \ln p(R))\,.
\label{p-voids-d}
\end{equation}
We expect that the above relation, with a suitable $A$, is also valid
if instead of taking a cube a size $2R$ we take a sufficiently compact
volume (e.g., a spheroid) linear size $R$. The mathematical treatment in
the case of isotropic displacements $p_d({\bf u})=p_d(u)$ is more
difficult, but we expect qualitatively the same result.  We give only
a rough sketch of this treatment.  Let us take a sphere of very large
radius $R$ and, as above, factorize the probability $W(R)$ that after
the application of the displacements it becomes empty into the product
of the probability $W_1(R)$ that all particles initially out of the
sphere stay out and the probability $W_2(R)$ that all the particles
initially in the sphere go out of it because of the displacements.  As
in the previous case, we expect that $W_2(R)$ is the dominating factor
for what concerns the large $R$ decreasing behavior of $W(R)$.  This
can be seen through the following arguments.  In order to evaluate
$W_1(R)$ at sufficiently large $R$, we approximate the probability
that a point, initially at a distance between $r$ and $r+\Delta r$
from the center of the sphere with $r\gg R$ and $\Delta r\ll r$, will
stay out of the sphere after the displacement, as
\[1-p_d(r) \frac{\Omega_d}{d}R^d\,,\]
with $\Omega_d$ the complete spherical angle in $d$ dimensions.  Now
the number of these particles in the initial lattice configuration is
around $n_0\Omega_d r^{d-1}\Delta r$. Therefore, by taking the product
over the spherical shells of thickness $\Delta r$ for radii greater
than $R$, we can write
\begin{equation}
W_1(R)\simeq \prod_{\Delta r} \left(1-p_d(r) 
\frac{\Omega_d}{d}R^d\right)^{n_0\Omega_d r^{d-1}\Delta r}\,.
\label{w1}
\end{equation} 
In the given limits Eq.~(\ref{w1}) can be reapproximated as
\begin{equation}
W_1(R)\simeq \exp \left[-n_0 \frac{\Omega_d^2}{d}R^d
\int_R^{+\infty}dr\,r^{d-1}p_d(r)\right]\,.
\label{w1-b}
\end{equation} 
In complete analogy with the one-dimensional case, it is simple now to
see that, if $p_d(u)$ decays faster than $u^{-2d}$ at large $u$, then
$W_1(R)$ at asymptotically large $L$ converges to a positive constant
$0<p_0<1$. Instead if $p_d(u)\sim B u^{-d-\beta}$ at large $u$ with
$0<\beta<d$ then
\[W_1(R)\simeq \exp\left[-aR^{d-\beta}\right]\,,\]
where the constant $a$ can be obtained approximatively by
Eq.~(\ref{w1-b}).  For what concerns the probability $W_2(R)$ we can
say that for sure it must be smaller than the probability $P(R)$ of
the following event: all the particles (whose number $N(R)$ is about
$\frac{\Omega_d}{d}(R/2)^d$) within a distance $R/2$ from the center
of the sphere make a displacement $u$ larger than $R/2$.  This
probability $P(R)$ is (in the large hypothesis Eq.~(\ref{tail}))
roughly given by
\[P(R)=\left(\Omega_d\int_{R/2}^{+\infty} du\,u^{d-1}p_d(u)\right)^{N(R)}
\simeq \exp\left[CR^d\ln p_d(R)\right]  \] 
with $C>0$ appropriate and depending on $d$, and where we have
considered the fact that, by definition, $p_d(u)$ decays faster
the $u^{-d}$ at large $u$.
On the other hand $W(R)$ must be larger than the 
probability $Q(R)$ that {\em all} the particles in the sphere make a jump of
size larger than $2R$. By similar reasoning one can find that
\[Q(R)\simeq \exp \left[DR^d\ln p_d(R)\right]\]
with $D$ another suitable constant depending on $d$.  Since $p_d(R)$
decrease to zero at large $R$, this shows at
the same time that the decaying behavior of the factor $W_2(R)$
prevails on the one of $W_1(R)$, and that again $W(R)$
must have the form given by Eq.~(\ref{p-voids-d}).

We recall now that in general in a Poisson point process in arbitrary
dimension and with average density $n_0$, the probability $W_P(V)$ that
a given volume $V$ is found empty of points is given by
\begin{equation}
W_P(V)= e^{-n_0V}\,.
\label{p-poisson}
\end{equation}
For point processes that are essentially Poisson with primarily positive and
short-range correlations, due to the only short-range
clusterization of points, we expect a similar relation for
sufficiently large voids, but with $n_0$ replaced by an appropriate
smaller constant \cite{torquato-book,To90}.  On the other hand,
in ``critical'' point processes, because of
the strong clusterization of points at all scales due to large-scale
positive correlations, we expect a larger probability of finding large
voids than in the Poisson one.  Therefore, for 
super-homogeneous point processes generated by displacing 
the points of a $d-$dimensional regular lattice
in an uncorrelated manner, the
probability that a compact volume of sufficiently large linear size $R$
decays with $R$ faster than in any non-super-homogeneous point process.

This observation suggests the following general heuristic conclusion:
\begin{itemize}
\item A point process is super-homogeneous {\em if and only if} its void size
distribution $W(R)$ satisfies the limit condition
\[\lim_{R\rightarrow +\infty} \frac{\ln W(R)}{R^d}=0.\]
\end{itemize}
Moreover, the above discussion about the void distribution generated in
a lattice by an uncorrelated but power-law-tailed displacement PDF
suggests a second general heuristic conclusion:
\begin{itemize}
\item A point process for which 
\[\lim_{R\rightarrow +\infty} \frac{\ln W(R)}{R^d\ln R}=0\]
not only is super-homogeneous but its power spectrum satisfies $s({\bf
k})\sim k^n$ at sufficiently small $k$ with $n\ge 2$.  However, this
proposition cannot be inverted. In fact, $n\ge 2$ is obtained also in
the case of power-law-tailed $p_d(u)$, but with a finite variance.

\end{itemize}

\section{Conclusions}

Super-Homogeneous stochastic point processes, and more generally
super-homogeneous mass stochastic density fields, are a very important
mathematical models of many systems not only in material science and
condensed matter physics, but also in diverse fields such as
cosmology.  For instance, slightly perturbed crystal lattices,
quasi-crystals \cite{quasi-cr}, one-component plasmas \cite{OCP1},
particular glassy systems, strictly jammed stochastic hard spheres
configurations \cite{TS2003} can all be seen as super-homogeneous point
density fields. Cosmological models predict a spectrum of the
primordial mass density perturbations of the Universe typical of
super-homogeneous systems \cite{CDM,OCP2}, and super-homogeneous point
processes (typically perturbed lattices or glass-like particle
distributions) are used as initial conditions in $n-$body simulations
to study the mass collapse and the structure (e.g, galaxies, and
clusters of galaxies) formation problems during the history
of the Universe.

Usually a point process is recognized to be super-homogeneous by
studying the scaling behavior of its number fluctuation $\sigma(R)$ with
respect to the distance at asymptotically large spatial scales (see
Eq.~(\ref{sup3bis})), or by analyzing the spatial integral of the
density covariance $C({\bf x})$ (see Eq.~(\ref{sup3})) or equivalently
the power spectrum $s({\bf k})$ at small wave numbers (see
Eq.~(\ref{sup3ter})).

In this paper, we have characterized super-homogeneous systems by
studying the statistical properties of the Voronoi cells and of void
size distribution. It is an important achievement because the
knowledge of the statistical properties of Voronoi cells is an
important issue in many subjects of disordered materials.  This task
has been accomplished mainly with the detailed study of the so-called
one-dimensional ``shuffled lattice'' i.e., a regular chain of
particles whose particle are randomly displaced from their lattice
positions with no correlations between the displacements.  Inspired by
the achievements obtained for these systems, we have generalized the
main results to the whole class of super-homogeneous point processes in
arbitrary spatial dimension.

The main results that we have obtained can be summarized as follows:
\begin{itemize}
\item For a particular subclass of one-dimensional ``shuffled
lattices'', one and two Voronoi cell statistics have been solved
exactly;
\item The correlation matrix $C_{ij}$ of the Voronoi cells of {\em
any} super-homogeneous point process satisfies a sum rule $\sum_j
C_{ij}=0$, which is independent of the way in which the single Voronoi
cells have been labeled. This is a very important relation because it
is a special property of only super-homogeneous point processes.
Indeed, this sum rule is the Voronoi-cell equivalent
of Eq.~(\ref{sup3}), which is the definition of a super-homogeneous
point process in terms of the covariance function.
\item In contrast to the conventional picture of super-homogeneous
systems, we have shown that arbitrarily large Voronoi cells or voids
are permitted in the super-homogeneous class. This is true despite the
fact that super-homogeneous point processes possess the slowest number
(mass) fluctuations-length scaling relation possible for any point
process.
\item For the most general one-dimensional shuffled lattice, 
we have found the asymptotic form of the void size distribution 
and its dependence on the ``shuffling'' statistics.
\item This result for void statistics has been extended to higher
dimensions, and suggests the introduction of two heuristic
conditions to identify and classify any
super-homogeneous point process in terms of the asymptotic behavior of
the void size distribution.  
\end{itemize}
This last result together with the sum rule about the correlation
matrix of the Voronoi cells are the two most significant achievements
of this study.  The present analysis and results open the possibility
for new studies on even more complex morphological characterizations of
super-homogeneous point processes.

\begin{acknowledgments}
\hspace{17.5pt}

A.G. thanks the Physics Department of the University ``La Sapienza'' 
of Rome (Italy) for having supported this research. 
S. T. gratefully acknowledges the support of the Office of Basic Energy Sciences, Department of Energy, 
under Grant No. DE-FG02-04ER46108.

\end{acknowledgments}

\appendix

\section{}
\label{app1}

In this appendix we give a brief derivation of Eq.~(\ref{disp1}).  For
a more general analysis of the effect of a stochastic displacement
field on the power spectrum of a given point process see
Ref. \cite{gabrielli1}.

Let us call $\hat n_I(x)=\sum_{i=1}^{N} \delta(x-x_i)$ the {\em
initial} microscopic density of a given point process, defined on
the line segment $[-L/2,l/2]$, where $-L/2\le x_i\le L/2$ is the
position of the $i^{th}$ point-particle of the system before the
application of the displacement field.  Ultimately, we will take the limit
$L\rightarrow\infty$. Let us also suppose
we know the global average density $n_0=\lim_{L_\rightarrow\infty} N/L$
and the power spectrum $s_I(k)$ of such a point process as defined
respectively by Eqs.~(\ref{vol-av}) and (\ref{ps-gen}).  We now apply
to each point $i$, independently of the
others, a stochastic displacement $u_i$ extracted from the
probability density function $p(u)$. The new microscopic density will be
\[\hat n(x)=\sum_i \delta(x-x_i-u_i)\,.\]
By definition, the new power spectrum $s(k)$ will be given by
\begin{equation}
s(k)=\lim_{L\rightarrow +\infty} \frac{1}{L}\left<
\overline{\sum_{i,j}^{1,N}e^{-ik(x_i-x_j+u_i-u_j)}}\right> 
-2\pi n_0^2\delta(k)\,,
\label{ap-eq1}
\end{equation}
where $\overline{(...)}$ stands for the average over all the possible
realizations of the displacement field for a given realization of the
{\em initial} point process, and $\left<...\right>$ stands for the
ensemble average over all the possible realizations of the initial
point process. In our hypothesis the displacement field and the point
process are considered statistically independent, and hence the two
averages commute and they can be taken in an arbitrary order.
We will first take the average over the displacements by separating
the diagonal contribution from the non-diagonal one in the double sum 
of Eq.~(\ref{ap-eq1}):
\[\overline{\sum_{i,j}^{1,N}e^{-ik(x_i-x_j+u_i-u_j)}}= N +
\left|\tilde p(k)\right|^2\sum_{i,j}^{1,N}\!\!\,' \,e^{-ik(x_i-x_j)}\,,\] 
where $\tilde p(k)$ is defined by Eq.~(\ref{char-f}) and $\sum_{i,j}'$
means the sum over all $i=1,...,N$ and $j=1,...,N$ with $i\ne j$.
Therefore, we can rewrite Eq.~(\ref{ap-eq1}) as
\begin{equation}
s(k)=\lim_{L\rightarrow +\infty}\left[ \frac{N}{L}\left(1-
\left|\tilde p(k)\right|^2\right)
+\frac{\left|\tilde p(k)\right|^2}{L}
\left<\sum_{i,j}^{1,N}e^{-ik(x_i-x_j)}\right>\right] - 2\pi n_0^2\delta(k)\,,
\label{ap-eq2}
\end{equation}
where we have added and subtracted the term 
$\frac{N}{L}\left|\tilde p(k)\right|^2=\sum_{i=1}^N 
\frac{\left|\tilde p(k)\right|^2}{L}$ in order to complete the double sum.
Equation (\ref{disp1}) is recovered by noticing that the following
relations hold:
\begin{eqnarray}
&&\tilde p(0)=1\nonumber\\
&&\lim_{L\rightarrow +\infty}\frac{N}{L}=n_0\nonumber\\
&&\lim_{L\rightarrow +\infty}\frac{1}{L}\left<\sum_{i,j}^{1,N}
e^{-ik(x_i-x_j)}\right> - 2\pi n_0^2\delta(k)=s_I(k)\,.\nonumber
\end{eqnarray}
The extension to higher spatial dimensions is obvious.


\begin{thebibliography}{99}
\bibitem{footnote1} The structure factor $S({\bf k}) \equiv 1+
n_0 {\tilde h}({\bf k})$ is proportional to the intensity
of scattered radiation from a system of point particles,
where ${\tilde h}({\bf k})$ is the Fourier transform of
total correlation function. Thus, the power spectrum $s({\bf k})$,
as we define it, is simply related to the structure factor by
the expression $s({\bf k})=n_0 S({\bf k})$.
\bibitem{footnote2}
The same  arguments can be simply
generalized to include asymmetric $p(u)$.
\bibitem{footnote3} It is simple to show that if we have an unlimited
tail that is faster than any generalized decreasing exponential (e.g.,
$p(R)\sim e^{-e^R}$), Eq.~(\ref{p-voids}) must be replaced by $W(R)\sim
\exp [AR \ln p(BR)]$ with $B>0$ of order $1$, which is
the general result for any unlimited tailed $p(u)$.
\end{thebibliography}
\end{document}